\documentclass[%
 aip,
 amsmath,amssymb,
 reprint,%
]{revtex4-1}

\usepackage{graphicx}  
\usepackage{dcolumn}   
\usepackage{bm}        
\usepackage{amssymb}   
\usepackage{amsmath}

\begin{document}

\title{Effects of tilt on the orientation dynamics of the large-scale circulation in turbulent Rayleigh-B{\'e}nard convection} 

\author{Dandan Ji}
\affiliation{Department of Physics, Yale University, New Haven, CT 06511, USA}

\author{Kunlun Bai}
\affiliation{Department of Mechanical Engineering and Materials Science, Yale University, New Haven, CT 06511, USA}

\author{Eric Brown}
\email{ericmichealbrown@gmail.com}
\affiliation{Department of Mechanical Engineering and Materials Science, Yale University, New Haven, CT 06511, USA}

Physics of Fluids 32 (7), 075118 (2020)

\date{\today}

\begin{abstract}
We experimentally test the effects of tilting a turbulent Rayleigh-B{\'e}nard convection cell on the dynamics of the large-scale circulation (LSC) orientation $\theta_0$.  The probability distribution of $\theta_0$ is measured, and used to obtain a tilt-induced potential acting on $\theta_0$, which is used in a low-dimensional model of diffusion of $\theta_0$ in a potential.  The form of the potential is sinusoidal in $\theta_0$, and linear in tilt angle for small tilt angles, which is explained by a simple geometric model of the vector direction of the mean buoyancy force acting on the LSC.  However,  the magnitude of the tilt-induced forcing 
is found to be two orders of magnitude larger than previously predicted.   When this parameter is adjusted to match values obtained from the probability distribution of $\theta_0$, the diffusive model can quantitatively predict effects of tilt on $\theta_0$.  In particular, tilt causes a change in potential barrier height between neighboring corners of a cubic cell, and changes in the barrier-crossing rate for $\theta_0$ to escape a corner are predicted with an accuracy of $\pm30\%$.   As a cylindrical cell is tilted, the tilt-induced potential provides a restoring force which induces oscillations when it exceeds the strength of damping; this critical tilt angle is predicted within 20\%, and the prediction is consistent with measured oscillation frequencies.  These observations show that a self-consistent low-dimensional model can be extended to include the dynamics of $\theta_0$ due to tilt.  However, the underprediction of the effect of tilt on $\theta_0$ warrants revisiting the predicted magnitude.
\end{abstract}



\maketitle

\section{Introduction}

Large-scale coherent flow structures are common in turbulence.  Examples of such structures include convection rolls in the atmosphere and oceans.  Such structures and their dynamics can play a significant role in heat and mass transport.  However, the Navier-Stokes equations that describe such flows are not practical to solve in many applications, so simpler models are desired.  Our long-term goal is to develop and test a general low dimensional model that can quantitatively predict the different dynamical states of large-scale coherent structures in different conditions.   In this manuscript, we extend an existing low-dimensional model \cite{BA08a, BA08b} by testing the effects of an additional forcing term due to tilting a convection cell relative to gravity on the dynamics of the orientation of convection rolls.

We  test the application of a low-dimensional model in the model system of turbulent Rayleigh-B\'enard convection.  A  fluid is heated from  below and cooled from above to generate buoyancy-driven flow  \cite{AGL09,LX10}. This system exhibits robust large-scale coherent structures that retain a similar organized flow structure over a long time. In particular, in containers of aspect ratio near 1, a large-scale circulation (LSC) forms. This LSC consists of localized blobs of coherent fluid known as plumes.  The plumes collectively form a single convection roll in a vertical plane  that can be identified in a turbulent background by averaging over the flow field for a short time \cite{KH81}.  
Turbulent fluctuations cause the LSC orientation $\theta_0$ in the horizontal plane to exhibit numerous dynamics.  For example, in level circular cylindrical containers, these dynamics include spontaneous and erratic meandering of the orientation $\theta_0$ \cite{BA06a}, cessation followed by reformation of the LSC \cite{BA06a, XX07, VDM19}, and oscillations of the structure \cite{HCL87, SWL89, CGHKLTWZZ89, CCL96, TSGS96, CCS97, QYT00, QT01b, NSSD01, QT02, QSTX04, FA04, SXT05,TMMS05, XZZCX09, ZXZSX09}.

To characterize the dynamics of the LSC, a low-dimensional model has been proposed, consisting of stochastic ordinary differential equations that describe diffusive motion in potentials \cite{BA07a, BA08a}.  The model terms are predicted by using the empirically known LSC structure as an approximate solution to the Navier-Stokes equations, and so the model can make predictions for the functional form and magnitude of terms.  The effects of turbulence are represented by stochastic terms.   This model and its extensions have successfully described most of the known dynamical modes of the LSC in level circular cylindrical  containers including the meandering, cessations \cite{BA07a,BA08a}, and oscillation modes described above \cite{BA09}, with the exception of the recently observed jump rope mode \cite{VHGA18}, which has not yet been modeled.  Predictions are generally quantitatively accurate within a factor of 3, but can be more accurate when more fit parameters are used \cite{AAG11}. 

 The low-dimensional model has been extended and tested with additional forcing terms to model other effects.  One example is the Coriolis force, which causes a  rotation of the LSC orientation \cite{BA06b,ZLW17, SLZ16}.  Another example is a potential dependent on the geometry of the cell, which predicts the tendency for the orientation $\theta_0$ to align with the longest diagonals, barrier-crossing as the LSC orientation escapes a corner of a cubic container to possibly reach a neighboring corner \cite{LE09, BJB16, FNAS17, GKKS18, VSFBFBR16, VFKSSV19}, and periodic oscillation between nearby adjacent corners of a rectangular-cross-section container \cite{SBHT14}.

Of particular interest here is the extension of the low-dimensional model to include tilting a convection cell relative to gravity \cite{BA08b}.  It was shown that an additional buoyant forcing in the thermal boundary layers due to tilt successfully predicts the increase in the strength of the LSC with tilt, and the resulting suppression of cessations \cite{BA08b}.  In a cylindrical cell, the buoyant forcing in the thermal boundary layers on $\theta_0$ due to tilt is predicted to lead to a locking of the orientation characterized by a narrower probability distribution $p(\theta_0)$ with increasing tilt angle. This trend has been observed qualitatively, but not tested quantitatively \cite{ABN06}.  At large tilt angles, a planar oscillation of $\theta_0$ around the orientation of the tilt was predicted as the tilt-induced forcing provides a restoring force, which was confirmed to exist \cite{BA08b}.  A rough consistency check was performed between the natural frequency and resonant frequency without accounting for damping \cite{BA08b}.  However, revisiting that test with more thorough analysis reveals an orders-of-magnitude discrepancy in the magnitude of the tilt-induced forcing that will be addressed in Sec.~\ref{sec:cylinder}.  Thus, the effect of tilt on the dynamics of the LSC orientation $\theta_0$ has never been tested quantitatively.  

In this manuscript, we quantitatively test how the dynamics of the LSC orientation $\theta_0$ depend on tilt.  We present new data from a cubic cell so we can test the effect of the tilt-induced potential on the barrier-crossing rate for the LSC orientation escaping a corner.  Experimental methods and the method to obtain the LSC orientation are explained in Sec.~\ref{sec:methods}.  The stochastic differential equation model is reviewed in Sec.~\ref{sec:model}.  Probability distributions of the LSC orientation $\theta_0$ at different tilt angles are shown in Sec.~\ref{sec:ptheta}.    The potentials are obtained from the probability distributions of $\theta_0$ in Sec.~\ref{sec:potential}.  The potentials are the most straightforward quantitative measurement of the orientation-dependent forcing due to tilt, and show that the effects of tilt on $\theta_0$ are orders of magnitude larger than predicted.  We use these potentials to predict modifications to the barrier crossing rate with tilt angle and test those predictions in Sec.~\ref{sec:switching}.   The magnitude of tilt effects is confirmed in a cylindrical cell, and the prediction of the frequency of a tilt-induced oscillation of $\theta_0$ around the orientation of the tilt angle is revisited in Sec.~\ref{sec:cylinder}.   Possible corrections to the model are discussed in Sec.~\ref{sec:discussion}.

\section{Methods}
\label{sec:methods}

\subsection{Experiment setup}

\begin{figure}
\includegraphics[width=0.2\textwidth]{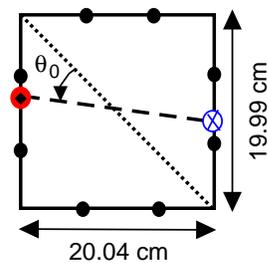} 
\caption{A cross section of the experimental setup viewed from above.  Thermistor locations on the sidewall are indicated by black circles.  The circulation plane of the LSC is indicated by the dashed line, where the hot  side is indicated by the circle with a diamond in the center, and the cold side is indicated by  the circle with a cross in the center.  The orientation of the LSC is defined as the angle $\theta_0$ between the hot side of the circulation plane and a longest diagonal line.  
}
\label{fig:setup}
\end{figure}   

The experimental apparatus is the same as was used in \cite{JB20a}, where it is described in detail. Briefly, the flow chamber consists of a nearly cubic cell of height $H=20.32$ cm and cross section  $20.04$ cm by $19.99$ cm.  The fluid in the chamber  is heated nearly uniformly from below and cooled from above by water pumped through aluminum plates, to hold a temperature difference $\Delta T$ between the top and bottom plates.  The sidewalls are plexiglas  to act as a thermal insulator.   To measure the LSC, thermistors were mounted in the sidewalls as in \cite{BA06b}.   We use measurements from one row of thermistors at the mid-height of the cell, equally spaced in the angle $\theta$ around the mid-plane, with $\theta=0$ defined at a corner, as illustrated in Fig.~\ref{fig:setup}.   Detailed information about the setup, thermistor calibration, and plate temperature uniformity is presented in \cite{JB20a}.
 
The working fluid is degassed and deionized water with mean temperature of 23.0$^{\circ}$C, for a Prandtl number $Pr = \nu/\kappa = 6.41$ where   $\nu=9.36\times10^{-7}$ m$^2$/s is the kinematic viscosity and $\kappa=1.46\times10^{-7}$ m$^2$/s is the thermal diffusivity. The Rayleigh number is given by $Ra= g\alpha \Delta T H^3/\kappa\nu$ where $g$ is the acceleration  of gravity, and $\alpha=0.000238$ K$^{-1}$ is the thermal expansion coefficient.  Unless otherwise specified, we report the measurement at $\Delta T=3.8$ $^{\circ}$C  for $Ra = 4.8\times10^8$.

\subsection{Method to obtain the tilt angle and its uncertainty}
\label{subsection:tilt_angle}

To control the tilt angle $\beta$, the apparatus is mounted on a base plate which is supported by three leveling screws.  One of the leveling screws at orientation $\theta_{\beta} = 3\pi/4$ rad (halfway between two corners) was turned (raised) to control the tilt angle $\beta$.  The tilt angle can be measured by a protractor.  Systematic errors were eliminated for each protractor measurement by averaging with the value obtained by reversing the orientation of the protractor.  After systematics were eliminated, the resolution of our protractor measurements was obtained as the standard deviation of repeated measurements to be $3\times10^{-4}$ rad.

However, we neglected to measure the tilt angle directly, instead calculated the tilt angle based on the number of revolutions of the leveling screws.  This introduced a random error on the tilt angle due to migration of the screws on the surface that they rested on, in addition to the resolution error of the protractor.  To estimate this extra error, we repeated a similar sequence of screw turns, this time measuring each tilt angle with a protractor. We found the error on the tilt angle to scale as a random diffusive process with the number of screw revolutions, and also depended on the number of times the screw was turned due to backlash.   We obtain the diffusivity $D_{\beta_y}= (7.3 \pm 3.9)\times 10^{-8}$ rad$^2$/rev from a fit of several measurements. The error on relative tilt angle between two measurements is estimated as $\Delta' \beta = \sqrt{D_{\beta_y}\zeta'}$,  where $\zeta'$ is the number of screw revolutions between measuring points (using the same step size as when temperature measurements were made).  The absolute error on tilt angles is calculated as  $\Delta \beta = \sqrt{D_{\beta_y}\zeta + (3\times10^{-4} rad)^2}$, where $\zeta$ is the number of screw revolutions from a reference point with a protractor measurement, and $3\times10^{-4}$ rad is the protractor resolution. The resulting absolute error $\Delta\beta$ is typically $8\times10^{-4}$ rad.

\subsection{Method to obtain the LSC orientation $\theta_0$} 
\label{sec:lsc}

To characterize the LSC in a noisy background of turbulent fluctuations, we fit the 8 thermistor temperatures at the mid-height of the cell to the function 
\begin{equation}
T = T_0 + \delta cos(\theta - \theta_0)
\label{eqn:delta}
\end{equation}
to obtain the orientation $\theta_0$ and strength $\delta$ of the LSC every 9.7 s, as in previous work \cite{BA06a, BA08a, BA08b, JB20a}.    
 An example of this fit and error analysis for this data is reported in \cite{JB20a}.   Systematic errors on $\theta_0$ from the calibration are 0.012 rad, although larger systematic effects exist due to the non-uniform temperature profile of the top and bottom plates \cite{JB20a}.


\section{Review of the low-dimensional model}
\label{sec:model}

\begin{figure}
\includegraphics[width=0.475\textwidth]{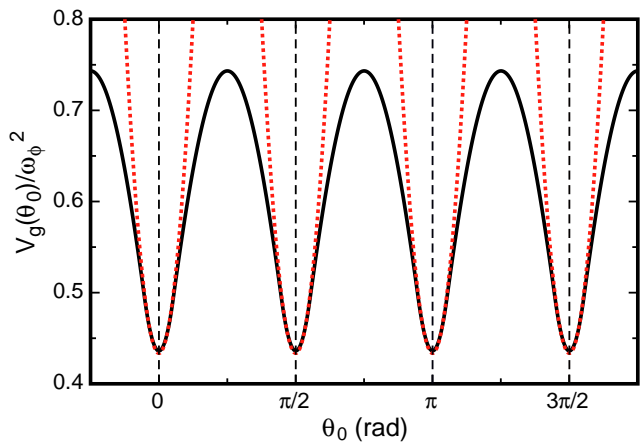}
\caption{Solid line: The model potential $V_g(\theta_0)$ for a cubic cell in the absence of tilt, reproduced from \cite{JB20a}.  The vertical dashed lines indicate the location of the four corners  where the potential minima are predicted. Equation \ref{eqn:ddottheta} describes diffusive fluctuations of $\theta_0$ in this potential.  Dotted lines: quadratic approximations of the potential around each corner.
}
\label{fig:potential}
\end{figure}

To model the dynamics of the LSC we use a previously derived stochastic ordinary differential equation \cite{BA08a}.  It uses the empirically known,  robust LSC structure as an approximate solution to the Navier-Stokes equations to obtain equations of motion for parameters that  describe the LSC dynamics.  The effects of fast, small-scale  turbulent fluctuations are separated from the slower, large-scale motion when  obtaining this approximate solution,  then added back in as a stochastic term in the ordinary differential equation.  The stochastic equation of motion for the LSC orientation $\theta_0$ in the horizontal plane in terms of the acceleration is
\begin{equation}
\ddot\theta_0 = -\frac{\delta\dot\theta_0}{\delta_0\tau_{\dot\theta}} -\nabla V(\theta_0) +f_{\dot\theta}(t) \ .
\label{eqn:ddottheta}
\end{equation}
The first term on the right side of the equation is a viscous damping term where  $\tau_{\dot{\theta}}$ is a damping time scale. The LSC strength $\delta$ fluctuates around its stable fixed point value, which is $\delta_0$ in an untilted cell and increases with tilt angle $\beta$ \cite{BA08b}.  While the fluctuations of $\delta$ can slightly modulate the damping, they do not have a significant effect on most of the dynamics of $\theta_0$ in this cell \cite{JB20a}, so we do not consider them here.  The potential $V(\theta_0)$ can have contributions from various forcing terms  \cite{BA08b}.  Here, we consider $V(\theta_0) = V_g(\theta_0) + V_{\beta}(\theta_0)$, where $V_g(\theta_0)$ is due to the container geometry, and $V_{\beta}(\theta_0)$ is due to tilting the cell relative to gravity.    $f_{\dot{\theta}}(t)$ is a stochastic forcing term which we model with a Gaussian distribution with diffusivity $D_{\dot\theta}$.   Equation \ref{eqn:ddottheta} is mathematically equivalent to diffusion in a potential landscape $V(\theta_0)$ with variable damping.

The geometric potential $V_g(\theta_0)$ is due to the pressure from the sidewalls, and was predicted as a function of the geometry of the container to be \cite{BA08b, BJB16}
\begin{equation}
V_g(\theta_0) = \left<\frac{ 3\omega_{\phi}^2 H^2 }{4 D(\theta_0)^2}\right>_{\gamma} \ .
\label{eqn:potential}
\end{equation} 
 $\omega_{\phi}$ is the angular turnover frequency of the LSC.   $D(\theta_0)$ is the distance across a horizontal cross section of the cell, for diagonal lines going through the centerpoint of the cell, as a function of $\theta_0$.    The  notation $\langle ...\rangle_{\gamma}$  represents a smoothing of the potential over an angular range of $\gamma=\pi/10$ in  $\theta_0$ due to the  non-zero width of the LSC \cite{SBHT14}.  This expression for $V_g(\theta)$ assumes fluctuations in the LSC strength $\delta$ are small, and consequences of those fluctuations are addressed in \cite{BJB16}.  The ratio $V_g(\theta_0)/\omega_{\phi}^2$ is a function of the geometry of the cell, and is shown in Fig.~\ref{fig:potential} for a cubic container.  The shape of $V_g(\theta_0)$ was found to qualitatively match predictions, with quantitative curvatures and barrier heights within a factor of 3 of predictions \cite{JB20a}.  

 The contribution of tilt to the potential $V_{\beta}(\theta_0)$ was derived from the buoyant forcing on the thermal boundary layers \cite{BA08b}.  The buoyant acceleration of a parcel of fluid in the thermal boundary layer parallel to the plate is $\dot u' = g\alpha(T-T_0)\sin\beta$.    About half the temperature drop in the sample occurs in each boundary layer, so the average temperature of each thermal boundary layer relative to the mean is approximately $T-T_0=\Delta T/4$.  This applies to a volume fraction of the fluid $2l/H$, where $l=H/2Nu$ is this thickness of a thermal boundary layer, and Nu is the Nusselt number, so the volume-averaged buoyant forcing on the LSC is $\langle\dot u'\rangle_V = g\alpha\Delta T\sin\beta/4Nu$.  A vector component of this acceleration  $\langle\dot u'\rangle_V\cos(\theta_0-\theta_\beta)$  is in the direction of the LSC and contributes to an enhancement of $\delta$ with tilt \cite{BA08b}, while another vector component with magnitude $-\langle\dot u'\rangle_V\sin(\theta_0-\theta_\beta)$ pushes the LSC horizontally to align it with the orientation $\theta_{\beta}$ of the tilt.  The latter acceleration is converted into an angular acceleration $\ddot\theta_0'$ of the LSC by taking a volume average of the acceleration divided by the radius, resulting in $\ddot\theta_0'= 3\langle\dot u'\rangle_V\sin(\theta_0-\theta_\beta)/H$.  A potential is obtained by integrating the forcing 
\begin{equation}
V_{\beta}(\theta_0) \equiv -\int \ddot\theta_0' d\theta_0  = - \omega_{\beta}^2 \sin\beta \cos(\theta_0-\theta_{\beta}) 
\label{eqn:potential_tilt}
\end{equation} 
where
\begin{equation}
\omega_{\beta}^2 = \frac{3\nu^2 Ra}  {4Nu Pr H^4} \ .
\label{eqn:omega_beta}
\end{equation}   
 Using the parameter values given in Sec.~\ref{sec:methods}, and the expected Nusselt number $Nu=48$ at this Ra$=2.8\times10^8$ based on fits of $Nu(Ra)$ to data from a cylindrical cell \cite{NBFA05, FBNA05}, the predicted value of $\omega_{\beta}^2=6.0\times 10^{-4}$ s$^{-2}$.

\subsection{Relation between the potential $V(\theta_0)$ and the probability distribution $p(\theta_0)$}

The potential $V(\theta_0)$ is related to a measured probability distribution $p(\theta_0)$ according to the steady state Fokker-Planck equation for Eq.~\ref{eqn:ddottheta} that describes the balance between forced advection and diffusion of probability density, assuming the overdamped limit where the term $\ddot\theta_0$ is negligible \cite{BA08b, SBHT14, JB20a}, and a negligible variation in $\delta$ so that it can be replaced by its mean value $\bar\delta$ which increases with $\beta$ \cite{BA08b}:
\begin{equation}
\frac{-\nabla V(\theta_0) \bar\delta}{\tau_{\dot\theta}\delta_0} p(\theta_0) = D_{\dot\theta} \frac{\partial p(\theta_0)}{\partial \theta_0} \ .
\label{eqn:fokkerplanck}
\end{equation}
 The solution is $p(\theta_0) = p_0\exp[-V(\theta_0)\bar\delta/D_{\dot\theta}\tau_{\dot\theta}\delta_0]$, where $p_0$ is a uniform factor determined by the normalization requirement that the integral of $p(\theta_0)$ equals 1.  We take the natural logarithm of both sides and rearrange this solution in the form of a dimensionless potential 
\begin{equation}
-\ln p(\theta_0)=\frac{V(\theta_0)\bar\delta}{D_{\dot\theta}\tau_{\dot\theta}\delta_0} - \ln p_0
\label{eqn:lnptheta}
\end{equation}

\section{The effect of tilt on probability distribution}
\label{sec:ptheta}
 
 \begin{figure}
 \centering
 \includegraphics[width=0.475\textwidth]{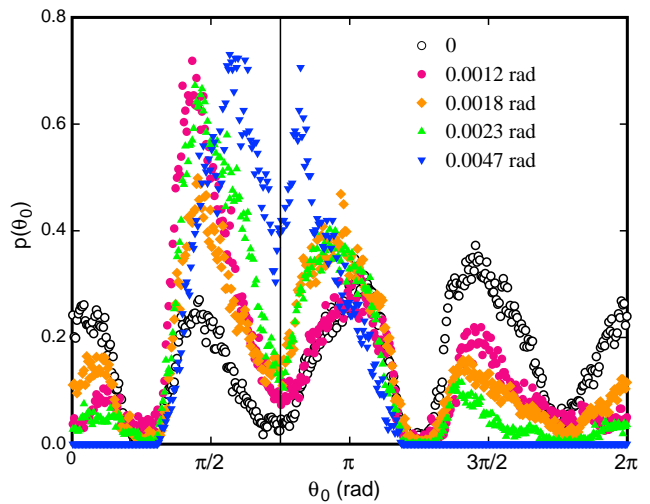}
 \caption{ Probability distributions $p(\theta_0)$ at various tilt angles $\beta'$ indicated in the legend. Vertical solid line: the tilt orientation $\theta_{\beta}$. As the tilt angle increases, $p(\theta_0)$ deviates from the more uniform distribution among four corners to favor the two corners closest to $\theta_{\beta}$. }
 \label{fig:pdf_theta0_VS_tilt_4C}
 \end{figure}

For a leveled cubic cell, 4 potential minima are predicted from Eq.~\ref{eqn:potential} as shown in Fig.~\ref{fig:potential}, one for each of the 4 corners of the cell.  According to Eq.~\ref{eqn:lnptheta}, this corresponds to 4 equal-sized peaks of $p(\theta_0)$.  However, we do not observe 4 equal-sized peaks in $p(\theta_0)$ in a nominally level cell -- instead we observe 1 to 3 peaks of different size, typically at the same locations, but with the number of peaks and the relative sizes depending on the duration of the experiment due to the long residence time in each potential well \cite{JB20a}.   The preference for some corners in $p(\theta_0)$ can primarily be attributed to the non-uniform temperature distribution in the top and bottom plates \cite{JB20a}, but may also be affected by other asymmetric forcings.    In an attempt to cancel out some of these asymmetries, we tilted the cell at different angles relative to gravity.  The most symmetric $p(\theta_0)$ we found, with 4 nearly-equal-sized peaks, was at a tilt angle of $\beta=0.0005 \pm 0.0009$ rad \cite{BJB16,JB20a}.   This $p(\theta_0)$ is reproduced in Fig.~\ref{fig:pdf_theta0_VS_tilt_4C}, reproduced from Ref.~\cite{JB20a}.  In this measurement, the LSC samples each potential well an average of 12.75 separate occasions over 6 days, still not entirely ergodic, as we estimate a relative uncertainty on peak sizes of $1/\sqrt{12.75}= 28\%$ based on Poisson statistics.   For ease of interpretation of tilt-induced effects at small tilt angles separate from other asymmetric forcings, we report results in terms of a shifted tilt angle $\beta'$ relative to this tilt angle where the probability distribution p$(\theta_0)$ is most symmetric.  

To quantify the effect of tilt on the probability distribution $p(\theta_0)$, we show $p(\theta_0)$ for several tilt angles $\beta'$ in Fig.~\ref{fig:pdf_theta0_VS_tilt_4C},  where the cell is tilted along the orientation $\theta_\beta = 3\pi/4$ rad (i.e.~in a plane parallel to sidewalls).  Errors on $\beta'$ relative to $\beta'=0$ are typically $5\times10^{-4}$ rad.   As the tilt angle $\beta'$ is increased from zero, Fig.~\ref{fig:pdf_theta0_VS_tilt_4C} shows a tendency away from a $p(\theta_0)$ with 4 nearly equal peaks to favor the 2 corners ($\theta_0=\pi/2$ rad and $\theta_0=\pi$ rad) on the side which was raised.  The probability distribution $p(\theta_0)$ was reduced to 2 peaks at $\beta' = 0.0047$ rad ($0.27^{\circ}$), identifying the small tilt angle required for tilt effects to overcome the geometric effects of the cubic shape of the cell.

 \section{Potential $V(\theta_0)$}   
\label{sec:potential}

\begin{figure}[]
\includegraphics[width=0.475\textwidth]{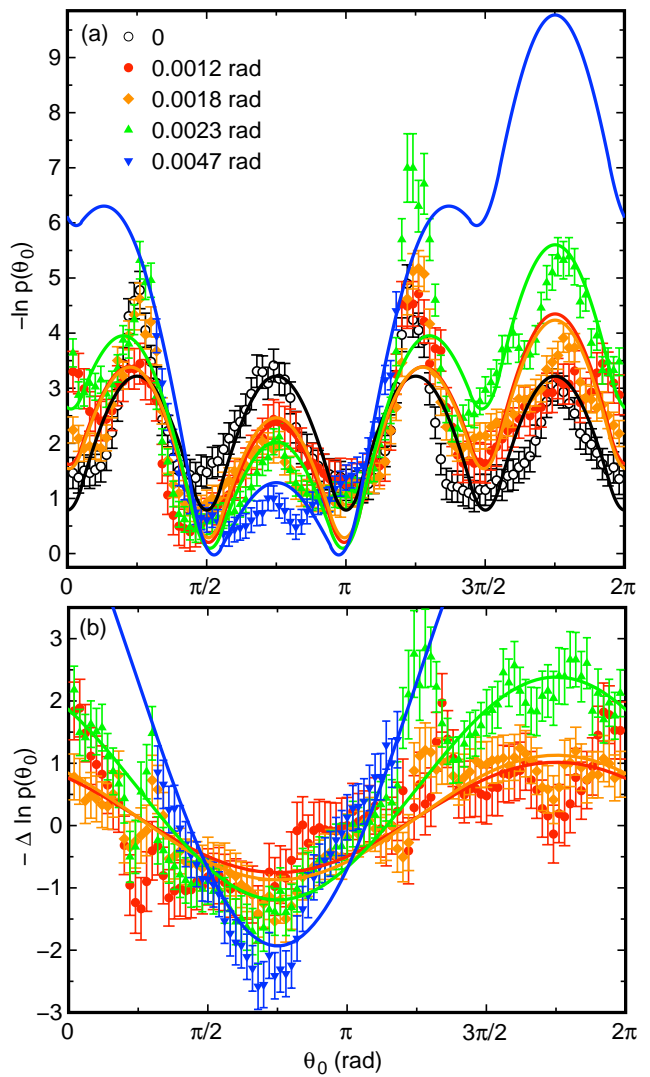}
\caption{(color online) (a) The dimensionless potential $-\ln p(\theta_0)$ for different tilt angles $\beta'$ indicated in the legend. (b) Difference $-\Delta\ln p(\theta_0)$ relative to the dimensionless geometric potential at $\beta'=0$ to obtain the tilt-dependent term of the potential.  Lines: fits of the model Eq.~\ref{eqn:Deltalnptheta} to obtain $\omega_{\beta}^2\bar\delta/D_{\dot\theta}\tau_{\dot\theta}\delta_0$.  The line colors match those of the corresponding data points.   The model dimensionless potential $V(\theta_0)/D_{\dot\theta}\tau_{\dot\theta}$ is reproduced as lines in panel a using values of $\omega_{\beta}^2\bar \delta/D_{\dot\theta}\tau_{\dot\theta} \delta_0$ from the fits in panel b.
}
\label{fig:lnptheta_cube}
\end{figure}  

We obtain the experimentally measured dimensionless potential  according to Eq.~\ref{eqn:lnptheta} by plotting $-\ln p(\theta_0)$.  This is shown in Fig.~\ref{fig:lnptheta_cube}a for the same tilt angles $\beta'$ as Fig.~\ref{fig:pdf_theta0_VS_tilt_4C}. The error bars represent the statistical error by adding in quadrature the error on each bin assuming Poisson statistics, with the error due to the Poisson sampling of each potential well as described in Sec.~\ref{sec:ptheta}. 

To test the prediction of the tilt contribution to $-\ln p(\theta_0)$ in the presence of a complex geometry-dependent term $V_g(\theta_0)$, we eliminate the geometry-dependent term by taking the difference in $-\ln p(\theta_0)$ between two tilt angles.  For a convenient reference point, we define $-\Delta\ln p(\theta_0)$ for a non-zero $\beta'$ as the difference in  $-\ln p(\theta_0)$ relative to the dimensionless potential at $\beta'=0$.  Assuming that $V_g(\theta_0)$ is independent of tilt, and inserting  Eq.~\ref{eqn:potential_tilt} into Eq.~\ref{eqn:lnptheta} yields
\begin{equation}
-\Delta\ln p(\theta_0) =  -\frac{\omega_{\beta}^2 \bar\delta\sin\beta' \cos(\theta_0-\theta_{\beta})}{D_{\dot\theta}\tau_{\dot\theta}\delta_0} - \Delta \ln p_0 
\label{eqn:Deltalnptheta}
\end{equation}
where the parameters $\bar\delta D_{\dot\theta}\tau_{\dot\theta}/\delta_0$ correspond to the value at the non-zero $\beta'$.  The measured difference $-\Delta\ln p(\theta_0)$ is shown in Fig.~\ref{fig:lnptheta_cube}b for each non-zero $\beta'$.  Fits of Eq.~\ref{eqn:Deltalnptheta} to the data are also shown, where $\omega_{\beta}^2\bar\delta/ D_{\dot\theta}\tau_{\dot\theta}\delta_0$ and $-\Delta \ln p_0$ are adjusted parameters. Since $-\Delta \ln p_0$ is determined by the normalization condition, it is not independent of $\omega_{\beta}^2\bar\delta/ D_{\dot\theta}\tau_{\dot\theta}\delta_0$, but we include it as a fit parameter to avoid having to fit with an implicit equation.  The fits have an average reduced $\chi^2=1.5$, indicating the predicted cosine shape is nearly as good a description of the data as possible given the size of the errors.

 Using the values of $\omega_{\beta}^2\bar \delta/D_{\dot\theta}\tau_{\dot\theta} \delta_0$ obtained for each tilt angle from fitting Eq.~\ref{eqn:Deltalnptheta},  we plot the corresponding  $-\ln p(\theta_0)$ from Eq.~\ref{eqn:lnptheta}  for each tilt angle in Fig.~\ref{fig:lnptheta_cube}a.  The data and model results both have the same general trends, confirming a self-consistency of the functional form of the model potential expressed by Eqs.~\ref{eqn:potential} and \ref{eqn:potential_tilt} and its relation to $p(\theta_0)$ according to the Fokker-Planck solution (Eq.~\ref{eqn:lnptheta}).  

\subsection{Obtaining a value for the magnitude of the tilt-induced forcing $\omega_{\beta}^2$}
\label{sec:omegabeta2}

\begin{figure}[]
\includegraphics[width=0.475\textwidth]{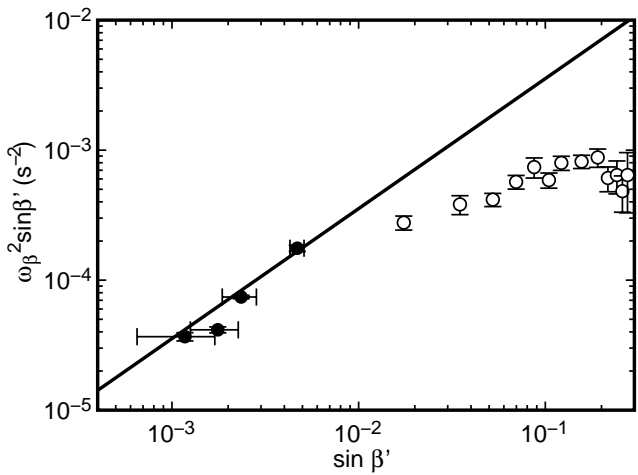}
\caption{Value of $\omega_{\beta}^2\sin\beta'$ as a function of $\sin\beta'$.  Solid circles: cubic cell at Ra $=4.8\times10^8$.  Open circles: cylindrical cell at Ra $=2.8\times10^8$. Line:  linear fit to cubic cell data to obtain $\omega_{\beta}^2$.  The data as small $\beta'$ is consistent with the linear trend in $\sin\beta'$ predicted by the model.  However, the value of the slope $\omega_{\beta}^2$ is nearly 2 orders of magnitude larger than predicted in both cells, and data at larger $\beta'$ no longer follow the predicted linear trend.
}
\label{fig:omega_beta2}
\end{figure}  

To calculate the value for the magnitude of the tilt-induced forcing $\omega_{\beta}^2$, we multiply the fit parameter $\omega_\beta^2\bar\delta/D_{\dot\theta}\tau_{\dot\theta} \delta_0$ obtained for each tilt angle in Fig.~\ref{fig:lnptheta_cube}b by $D_{\dot\theta}\tau_{\dot\theta}\delta_0/\bar\delta$.  Rather than characterize the tilt-dependence of each parameter separately,  $D_{\dot\theta}\tau_{\dot\theta}\delta_0/\bar\delta$ is measured altogether as the plateau of the mean-squared displacement of $\dot\theta_0$ over time, which is an expected relation from Eq.~\ref{eqn:ddottheta} in the limit of fixed damping and no potential \cite{BA08b}.   To confirm that these are valid assumptions, we carried out numerical solutions of Eq.~\ref{eqn:ddottheta} and the corresponding stochastic equation of motion for $\delta$ \cite{BA08b}, from which we found the approximation overestimates the plateau value $D_{\dot\theta}\tau_{\dot\theta}\delta_0/\bar\delta$ by typically a few percent, and generally within reported errors in $\omega_{\beta}^2$.  For this data series at small tilt angle $\beta' \le 0.0047$ rad, we measure a standard deviation in $D_{\dot\theta}\tau_{\dot\theta}\delta_0/\bar\delta$ of only 3.6\%, so we also use the approximation that this ratio of parameters is constant for this data series. We use the value $D_{\dot\theta}\tau_{\dot\theta}\delta_0/\bar\delta = 4.15\times 10^{-5}$ rad$^2$/s$^2$ from a different dataset nominally at $\beta'=0$ \cite{BJB16}. 

The resulting magnitude of the tilt-dependent forcing $\omega_{\beta}^2\sin\beta'$ is shown as a function of $\sin\beta'$ in Fig.~\ref{fig:omega_beta2}.  The errors on $\omega_{\beta}^2\sin\beta'$ represent the uncertainty from the fit of Eq.~\ref{eqn:Deltalnptheta}.  
The errors shown on $\sin \beta'$ are the errors $\Delta' \beta$ relative to $\beta' = 0$, which is the dominant error on the data. 
A linear fit to the cubic cell data for $\beta' \le 0.0047$ rad yields a value of $\omega_{\beta}^2=0.035\pm0.003$ s$^{-2}$ with reduced $\chi^2=0.7$.  The linear fit is consistent with a constant $\omega_{\beta}^2$ as predicted. However, the value is 1.7 orders of magnitude larger than the predicted value of $\omega_{\beta}^2$ \cite{BA08b},  indicating  a quantitative failure of the model much larger than the typical prediction error of a factor of 2 or 3 \cite{BA08a}.

 \section{Barrier crossing}
\label{sec:switching}
\subsection{Time series of $\theta_0$}

 \begin{figure}
 \centering
 \includegraphics[width=0.475\textwidth]{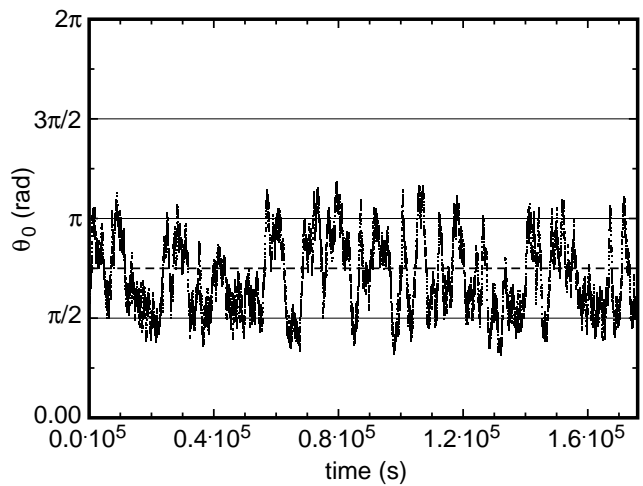}
 \caption{Time series of the LSC orientation $\theta_0$ at tilt angle $\beta'= 0.0047$ rad.  Corners are indicated by horizontal solid lines.  $\theta_0$ clusters around the two corners closest to the tilt orientation $\theta_{\beta}=3\pi/4$ (horizontal dashed line), and occasionally crosses the potential barrier at $\theta_0=3\pi/4$ to move between corners}
 \label{fig:theta_time}
 \end{figure}
 
 A time series of the LSC orientation $\theta_0$ is shown for a tilt angle $\beta'= 0.0047\pm0.0008$ rad in Fig.~\ref{fig:theta_time}.  The time series is dominated by fluctuations on the short term.  On the long term, data clusters around the two preferred corners at $\theta_0 = \pi/2$ and $\pi$ with lower potentials as shown in Fig.~\ref{fig:lnptheta_cube}a.    The LSC orientation $\theta_0$ only occasionally switches between these corners, indicating a significant potential barrier between the corners.   This time series is similar to the time series in Ref.~\cite{BJB16}, although in that more leveled cell, $\theta_0$ sampled and clustered around all four corners.   Because tilting the cell changes the potential barriers between corners as shown in Fig.~\ref{fig:lnptheta_cube}a, tilt is expected to have a significant effect on the rates of barrier crossing between different corners.

\subsection{Prediction of barrier crossing rates $\omega_{ij}$}

\begin{figure}[]
\includegraphics[width=0.475\textwidth]{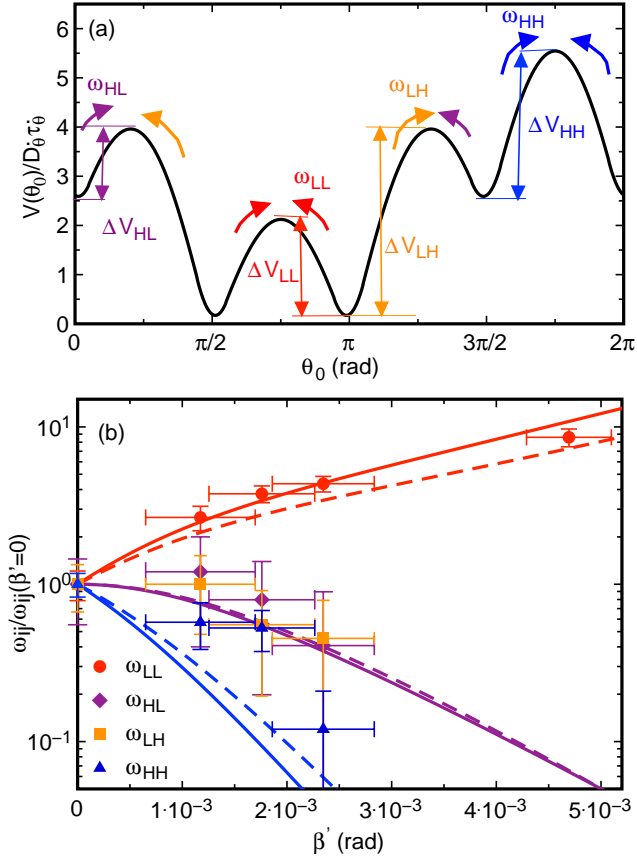}
\caption{(color online) (a) Schematic of the dimensionless potential for $\beta' > 0$ and $\theta_{\beta}=3\pi/4$, illustrating the different barrier crossing rates $\omega_{ij}$ and corresponding potential barriers $\Delta V_{ij}$.  (b) Solid symbols: The measured barrier crossing rate $\omega_{ij}$  normalized by the values in the untilted cell $\omega_{ij}(\beta'=0)$,  where the indices $i$ and $j$ correspond to different potential wells and barriers as indicated in the legend.  Solid lines: the model results using independently measured values for model parameters. Dashed line: the model results using the measured value for $\omega_{\beta}^2$ and predicted values for parameters relating to effects of cell geometry.  The same color scheme is used for the data and predictions, except for $\omega_{LH}$ which is predicted to have the same values as  $\omega_{HL}$.   The model captures the trends well, showing that the effects of tilt on the orientation of the LSC can be predicted by a low-dimensional model.
 }
\label{fig:omega_beta}
\end{figure}

  In this section we use the tilt parameter $\omega_{\beta}^2$ obtained from Sec.~\ref{sec:omegabeta2} and $V_g(\theta_0)$ from Eq.~\ref{eqn:potential} with parameter values from \cite{JB20a} to predict the frequency of the LSC orientation escaping a corner.  This barrier crossing is described by the Kramers model \cite{Kr40}.  The model prediction is an extension of our previous work in an untilted cubic cell \cite{BJB16}.    Tilting the cell at $\theta_{\beta} =3\pi/4$ rad creates a lower potential at at the nearby corners, while it creates a higher potential at the farther corners as was seen in Fig.~\ref{fig:lnptheta_cube}.   Barrier crossing rates $\omega_{ij}$ and corresponding potential barriers $\Delta V_{ij}$ from well $i$ to well $j$ are labeled in an illustration of the potential in Fig.~\ref{fig:omega_beta}a, where $i$ and $j$ can each stand for either the lower potential ($L$) or higher potential ($H$).  With tilt at $\theta_{\beta}=3\pi/4$ rad, there are two unique potential minima values, and so four unique barrier crossing rates $\omega_{ij}$ and potential barriers $\Delta V_{ij}$.

 The rate at which $\theta_0$ crosses a potential barrier $\Delta V$ is estimated from the Kramers model, in the overdamped limit of Eq.~\ref{eqn:ddottheta}, and assuming $\delta=\delta_0$ is a constant \cite{BJB16}:
\begin{equation}
\omega_{ij} =  p_{i}\frac{ \sqrt{c_{min,i}c_{max,ij}}\tau_{\dot{\theta}} }{2\pi} \exp \left( - \frac{ \Delta V_{ij}}{D_{\dot{\theta}} \tau_{\dot{\theta}}} \right) \ .
\label{eqn:switching_rate}
\end{equation}   
 $c_{min,i}$ is the curvature of the potential at the local minimum of well $i$ and  $c_{max,ij}$ is the curvature of the potential at the local  maximum of the barrier between wells $i$ and $j$.  The factor $p_i$ corresponds to the probability of the LSC being in well $i$, so that $\omega_{ij}$ corresponds to the rate of barrier crossing averaged over a time series where each well is sampled ergodically and according to their potentials.  

The potential curvatures $c_{min,i}$, $c_{max,ij}$, and the potential barrier $\Delta V_{ij}$ are related to model parameters by performing quadratic expansions  in $\theta_0$ around each local minimum and maximum of the potential from Eq.~\ref{eqn:potential_tilt}, respectively, as $V_i \approx  V_{min,i} + (c_{min,i}/2)(\theta_0-\theta_{min,i})^2$ and $V_{ij}\approx V_{max,ij} - (c_{max,ij}/2)(\theta_0-\theta_{max,ij})^2$.  The potential barrier is defined as $\Delta V_{ij} \equiv V_{max,ij}-V_{min,i}$.   Using the labels $c_{max}$,  $c_{min}$, and $\Delta V_g $ to refer to the values for the untilted cell  where the four wells are nominally identical \cite{BJB16}, the special case of the tilt orientation $\theta_{\beta}=3\pi/4$ rad being halfway in between two corners results in the terms

\begin{align}
\label{eqn:cmin}
c_{max,LL} &= c_{max} - \omega_{\beta}^2\sin\beta\\
c_{max,HH} &= c_{max} + \omega_{\beta}^2\sin\beta\\
c_{max,LH}  &= c_{max,HL} = c_{max}\\
c_{min,L} &= c_{min} + (\omega_{\beta}^2\sin\beta)/\sqrt{2}\\
c_{min,H} &= c_{min} - (\omega_{\beta}^2\sin\beta)/\sqrt{2} 
\end{align}
\begin{multline}
\Delta V_{LL} = \Delta V_g -\\ 
\left(1-\frac{1}{\sqrt{2}}+\frac{\omega_{\beta}^2\sin\beta}{4\left(c_{max,LL}+\frac{\omega_{\beta}^2\sin\beta}{\sqrt{2}}\right)}\right)\omega_{\beta}^2\sin\beta
\end{multline}
\begin{multline}
\Delta V_{HH} = \Delta V_g +\\ 
\left(1-\frac{1}{\sqrt{2}}-\frac{\omega_{\beta}^2\sin\beta}{4\left(c_{max,HH}+\frac{\omega_{\beta}^2\sin\beta}{\sqrt{2}}\right)}\right)\omega_{\beta}^2\sin\beta
\end{multline}
\begin{multline}
\Delta V_{LH} = \Delta V_g + \\
\left(\frac{1}{\sqrt{2}}+\frac{\omega_{\beta}^2\sin\beta}{2c_{max,LH}}+\frac{\omega_{\beta}^2\sin\beta}{4\left(c_{max,LH}+\frac{\omega_{\beta}^2\sin\beta}{\sqrt{2}}\right)}\right)\omega_{\beta}^2\sin\beta
\end{multline}
\begin{multline}
\Delta V_{HL} = \Delta V_g - \\
\left(\frac{1}{\sqrt{2}}-\frac{\omega_{\beta}^2\sin\beta}{2c_{max,HL}}+\frac{\omega_{\beta}^2\sin\beta}{4\left(c_{max,HL}+\frac{\omega_{\beta}^2\sin\beta}{\sqrt{2}}\right)}\right)\omega_{\beta}^2\sin\beta \ .
\label{eqn:DeltaV}
\end{multline}

To obtain the probability $p_i$ of being in a well, we integrate over the probability distribution obtained from the Fokker-Planck equation (Eq.~\ref{eqn:fokkerplanck}) with the approximation $\bar\delta=\delta_0$.  To perform the integral, we apply a 2nd order Taylor expansion in $\theta_0$ and use the same trick as Kramers \cite{Kr40} to approximate the integral over a Gaussian distribution in one well as the integral over a Gaussian from $-\infty\rightarrow\infty$ so $p_i \propto \sqrt{2\pi/c_{min,i}}\exp(-V_{min,i}/D_{\dot\theta}\tau_{\dot\theta})$.  The proportionality is eliminated by using the normalization requirement $p_L+p_H=1$, and substituting $V_{min,L}-V_{min,H} = \Delta V_{HL} - \Delta V_{LH}$ yields 
\begin{equation}
p_L = \left[1+\sqrt{\frac{c_{min,L}}{c_{min,H}}}\exp\left(\frac{\Delta V_{HL} - \Delta V_{LH}}{D_{\dot\theta}\tau_{\dot\theta}}\right)\right]^{-1} \ .
\label{eqn:prob_i}
\end{equation}
\noindent   $p_H$ is given by a similar expression with $H$ and $L$ interchanged. 

Predictions of the barrier crossing rate $\omega_{ij}$ as a function of $\beta$ are shown in Fig.~\ref{fig:omega_beta}b for the different barriers using Eqs.~\ref{eqn:cmin} -- \ref{eqn:prob_i} for the terms in Eq.~\ref{eqn:switching_rate}. To focus on the tilt-dependence for the different barriers, curves are normalized by the barrier-crossing rate $\omega_{ij}$ at $\beta=0$.   We use the measured parameter values at $\beta'=0$  of $c_{min}=2.3\times10^{-4}$ s$^{-2}$ and $c_{max}=2.2\times10^{-3}$ s$^{-2}$ from Ref.~ \cite{BJB16},  $\Delta V_g/D_{\dot\theta}\tau_{\dot\theta} = 2.61$ from Ref.~\cite{JB20a}, and $\omega_{\beta}^2=0.035$ s$^{-2}$ from Sec.~\ref{sec:omegabeta2}.  

The trends in predicted barrier crossing rates $\omega_{ij}$ with $\beta$ in Fig.~\ref{fig:omega_beta}b are explained as follows.  $\omega_{LL}$ increases with tilt angle because the potential barrier $\Delta V_{LL}$ decreases as seen in Fig.~\ref{fig:omega_beta}a, and the probability $p_L$ of being in the lower potential well increases.  In contrast, $\omega_{HH}$  decreases with tilt angle because $\Delta V_{HH}$ increases and $p_H$ decreases.  $\omega_{LH}$ and $\omega_{HL}$ are intermediate because the change in potential barrier and probability of being in the well have opposite trends.   Note that the predictions for $\omega_{LH}$ and $\omega_{HL}$ are equivalent, as is required for a closed system with no net flux between high and low potential states. 

The most significant change to the barrier-crossing rate $\omega_{ij}$ with $\beta$ at small $\beta$ are the terms in $\Delta V_{ij}$ that are linear in $\sin\beta$, which change $\Delta V_{ij}$ by up to 40\% over the range shown in Fig.~\ref{fig:omega_beta}b, which is further amplified by the exponential term in Kramers' equation (Eq.~\ref{eqn:switching_rate}) to be up to a factor of 3 change in $\omega_{ij}$.  The remaining terms in Eqs.~\ref{eqn:cmin} -- \ref{eqn:DeltaV} are small for $\beta \le 0.0047$ rad, since $\omega_{\beta}^2\sin\beta$ is small compared to $c_{min}$ and $c_{max}$ (this can be seen by comparing the curvatures at the minimum in Fig.~\ref{fig:lnptheta_cube}b to the curvatures in Fig.~\ref{fig:lnptheta_cube}a at $\beta=0$).  The changes to $c_{min,i}$ with $\beta$ are at most 20\%, while the changes in $c_{max,ij}$ are at most 3\% for $\beta \le 0.0047$ rad.  These effects are halved in the calculation of $\omega_{ij}$.   Similarly, the terms in $\Delta V_{ij}$ that are 2nd order in $\sin\beta$ cause at most a 1\% correction on $\omega_{ij}$.   At larger $\beta$, not only would these terms increase in significance, but the increase in $\bar\delta$ with tilt would also be more significant, which can be straightforwardly included in the prediction of $\omega_{ij}$ \cite{BA08b}.

\subsection{Measurement of barrier crossing rates}

To compare with Kramers' prediction of the barrier crossing rate, we measure the number of barrier crossing events when the LSC escapes a stable potential minimum to reach a potential maximum. To avoid counting the jitter around the potential maximum as multiple crossing events, we count an event when $\theta_0$ crosses the potential maximum, but do not count another event  again until $\theta_0$ first reaches near the stable fixed point (within $\pi/8$ of a corner).  This is the same definition as what we used in \cite{BJB16}. 
Using this this definition, it is possible to count barrier crossing events where the LSC orientation falls back into the same potential well it was in before it reached the potential maximum, or into the potential well of an adjacent corner.   The barrier crossing rate $\omega_{ij}$ is measured as the count of barrier crossings divided by the overall duration of the time series. 

To isolate the effect of tilt on the barrier crossing rates, measured values of $\omega_{ij}$ are normalized by  the barrier crossing rate at zero tilt $\omega_{ij}(\beta'=0)$.  This normalization is done separately for each barrier to remove the effects of weak  unintended asymmetries of the setup, mostly due to the nonuniformity of the top and bottom plate temperatures \cite{JB20a}.  These asymmetries lead to differences in the potential barriers $\Delta V_{ij}$ observed at $\beta'=0$  in Fig.~\ref{fig:lnptheta_cube}a and thus different barrier crossing rates for the different barriers at $\beta'=0$.  Such a  barrier-dependent additive term in the potential  would lead to a multiplicative factor in the barrier crossing rate in Eq.~\ref{eqn:switching_rate}, so the effects of tilt can be isolated from other barrier-dependent effects by dividing measured values of $\omega_{ij}$  by  the barrier crossing rate at zero tilt $\omega_{ij}(\beta'=0)$ separately for each barrier.  

Figure \ref{fig:omega_beta}b shows the normalized measured barrier-crossing rate $\omega_{ij}$ as a function of tilt angle $\beta'$.  Random errors on $\beta'$ are reported as the uncertainty $\Delta'\beta$ relative to $\beta'$ where the data is normalized.  Fractional errors on $\omega_{ij}$ are 
reported as one over the square root of the count of events assuming Poisson statistics.  For $\beta'>0$, the error on the normalized $\omega_{ij}$ includes the fractional error on $\omega_{ij}(\beta'=0)$ added in quadrature with that of $\omega_{ij}$.

The general trends in Fig.~\ref{fig:omega_beta}b are similar for both the measurements and the predictions, although the prediction for $\omega_{LL}$ appears to systematically underestimate the barrier crossing rates. On average the predictions vary from the data by a difference of 30\%, well within the typical accuracy of this model of about a factor of 3 due to approximations made about the shape of the LSC, scale separation between the LSC and small-scale turbulent fluctuations, and the distribution of turbulent fluctuations \cite{BA08a, BA08b, BA09, SBHT14, BJB16}.  Statistically, we use a chi-squared test, calculating the reduced $\chi^2$ as the mean-squared difference between the measured and predicted $\omega_{ij}/\omega_{ij}(\beta'=0)$, divided by the square of the error on the normalized $\omega_{ij}$ (we do not include the errors on $\beta'$ in this calculation).   We calculated the reduced $\chi^2$ for predictions with different values of $\omega_{\beta}^2$ and found the range of $\omega_{\beta}^2=0.037\pm0.008$ s$^{-2}$ results in predictions consistent with the measured normalized values of $\omega_{ij}$ with a reduced $\chi^2<1.7$, corresponding to a 95\% confidence interval for 13 data points.  This best fit range is consistent with the value of $\omega_{\beta}^2 = 0.035\pm0.003$ s$^{-2}$ obtained from fitting $p(\theta_0)$, confirming the self-consistency of the stochastic model of Eq.~\ref{eqn:ddottheta} with the potentials characterized by Eq.~\ref{eqn:potential} and \ref{eqn:potential_tilt}.


At our largest reported tilt angle of  $\beta' = 4.7\times10^{-3}$ rad, $\omega_{LH}$, $\omega_{HL}$  and $\omega_{HH}$ are measured to be 0, as the rate drops below our resolution limit of 1 event over the duration of the experiment.  In terms of the normalized barrier crossing rate, this is a resolution of 0.044 for $\omega_{HH}$,  0.17 for $\omega_{LH}$, and 0.30 for $\omega_{HL}$, which differ because of the different measured barrier crossing rates at $\beta'=0$.  Since the prediction drops below these resolution limits, then the lack of measured barrier crossing at this tilt angle is consistent with the prediction.

\subsection{Effects of different model assumptions}

Variations in $\delta$ are known to have a significant effect on the barrier crossing rate as the effective potential barrier and damping both decrease as $\delta$ decreases, corresponding to a multiplicative factor of $\delta_0/\delta$  and a factor of $(\delta/\delta_0)^3$ in the exponent of Eq.~\ref{eqn:switching_rate} for the barrier-crossing rate $\omega_{ij}$ \cite{BJB16}.   The mean value $\bar\delta$ does not vary by more than 1.8\% for $\beta' \le 0.0047$. This would only lead to a correction of up to 13\% on $\omega_{ij}$ at large $\beta'$, 
which is not included in the presented model because it is small compared to the other errors on the data in Fig.~\ref{fig:omega_beta}b.  Furthermore, the standard deviation of $\delta/\delta_0$ does not vary with $\beta'$ for $\beta' \le 0.0047$ rad, within errors of 6\%.   This suggests that any correction to the total barrier crossing rate  due to fluctuations in $\delta$ \cite{BJB16} would be the same at different tilt angles, and would cancel out with the normalization by $\omega_{ij}(\beta'=0)$.  Thus, these calculations confirm that assuming $\delta=\delta_0$ in Eq.~\ref{eqn:potential} is a valid approximation for this data.  

 We also consider calculating the barrier crossing rate $\omega_{ij}$ using predictions instead of experimental measurements for the geometric component of the model. 
 If we calculate $\omega_{ij}$ using predictions $c_{max} = 3\omega_{\phi}^2/2$,  $c_{min} = 15\omega_{\phi}^2/\pi$, and $\Delta V_g = (3/8)(1-\gamma/2)\omega_{\phi}^2$ \cite{BJB16}, and the turnover rate $\omega_{\phi}=0.022$ rad/s$^{-1}$ measured at $\beta'=0$ at the same Ra and consistent with the Grossman-Lohse model prediction \cite{BJB16}, we obtain the dashed lines in Fig.~\ref{fig:omega_beta}b.  These are close to the predictions using the measured values for the geometric parameters, and consistent with the data (reduced $\chi^2=1.3$) which reinforces the  conclusion that the contribution of the geometry to the dynamics of the LSC can be predicted by the model \cite{BJB16}.

\section{$\omega_{\beta}^2$ in a cylindrical cell and larger tilt angles}
\label{sec:cylinder}

In Secs.~\ref{sec:omegabeta2} and \ref{sec:switching}, we found that the tilt-induced-forcing parameter $\omega_{\beta}^2$ was 1.7 orders of magnitude larger than predicted \cite{BA08b}.  To confirm whether the value of $\omega_{\beta}^2$ is universally larger than predicted, or if this is only the case for a cubic cell, we also obtain values of $\omega_{\beta}^2$ from a cylindrical cell.  Data is taken from a previous experiment at $\Delta T=0.5$ K for Ra $=2.8\times10^8$ and Pr $=4.4$ with tilt angles up to $\beta'=0.28$ rad \cite{BA08b} (the difference between $\beta$ and $\beta'$ is negligible for this dataset due to the larger tilt angles measured).    We follow the same procedure of fitting $-\Delta\ln p(\theta_0)$ at different tilt angles to obtain $\omega_{\beta}^2 \bar\delta/D_{\dot\theta}\tau_{\dot\theta} \delta_0 $ as in Sec.~\ref{sec:potential}.  
Since this dataset is over a wider range of $\beta'$, the variation of model parameters with $\beta'$ becomes more significant.  Thus, for this data series, we obtain the normalization factor $D_{\dot\theta}\tau_{\dot\theta} \delta_0/\bar\delta$ separately for each value of $\beta'$ as the plateau of the mean-squared displacement of $\dot\theta_0$ at large time intervals \cite{BA08b}. This value  $D_{\dot\theta}\tau_{\dot\theta} \delta_0/\bar\delta$ decreases by a factor of 7.7 as $\beta'$ is increases to 0.28 rad.  This variation can be further broken down into variations of the individual parameters.  $\bar\delta$ increases with $\beta'$ by a factor of 2.2 over this range due to the strengthening of the LSC strength along the orientation of tilt \cite{BA08b}.  $D_{\dot\theta}$ is obtained from the initial slope of the mean-squared displacement of $\dot\theta_0$ over time \cite{BA08a}, and is found to decrease by a factor of 5.5 over this range of $\beta'$.  A trend in this direction is expected if the temperature fluctuation strength is independent of $\beta'$; the larger $\bar\delta(\beta')$ at larger tilt angles corresponds to a larger angular momentum of the LSC, which makes it harder for the LSC to reorient, corresponding to a decrease in diffusivity $D_{\dot\theta}$.   

The resulting values of $\omega_{\beta}^2\sin\beta'$ from the cylindrical cell data are shown in Fig.~\ref{fig:omega_beta2}.  They indicate an initially increasing trend in $\sin\beta'$, however the forcing appears to level off at large $\beta' >0.1$ rad, in disagreement with the prediction of a linear scaling in $\sin\beta'$.  Corresponding values of $\omega_{\beta}^2$ range from 0.016 s$^{-2}$ at the smallest $\beta'$ to $0.0023$ s$^{-2}$ at the largest $\beta'$, from 1 to 2 orders of magnitude larger than the predicted $\omega_{\beta}^2 = 1.3\times10^{-4}$ s$^{-2}$ based on Eq.~\ref{eqn:omega_beta} and measured parameter values from \cite{FBNA05,BA08b}.  At the lowest $\beta'$, this discrepancy is similar to that found in the cubic cell at small $\beta'$ in Sec.~\ref{sec:potential}, suggesting that the effect of tilt is similar for the two cell geometries, and the disagreement with the predicted value of $\omega_{\beta}^2$ originates from something other than cell geometry.

\subsection{Revisiting tilt-induced oscillations}

Only one feature of the system dependent on the value of $\omega_{\beta}^2$ was tested previously -- a planar oscillation of the LSC around the tilt orientation $\theta_{\beta}$ induced by a large tilt of the cell \cite{BA08b}.  The oscillation frequency was assumed to be near the natural frequency of the potential (i.e.~ignoring the damping term in Eq.~\ref{eqn:ddottheta} for a harmonic oscillator), corresponding to $\omega_{\beta}^2 = 3.6\times10^{-4}$ s$^{-2}$ for the same data shown in Sec.~\ref{sec:cylinder}, only 2.8 times the prediction \cite{BA08b}.  Since an alternate method of obtaining $\omega_{\beta}^2$ was not tested at the time\cite{BA08b}, this was deemed a sufficient consistency and was not investigated further with a more complex model.

\begin{figure}[]
\includegraphics[width=0.475\textwidth]{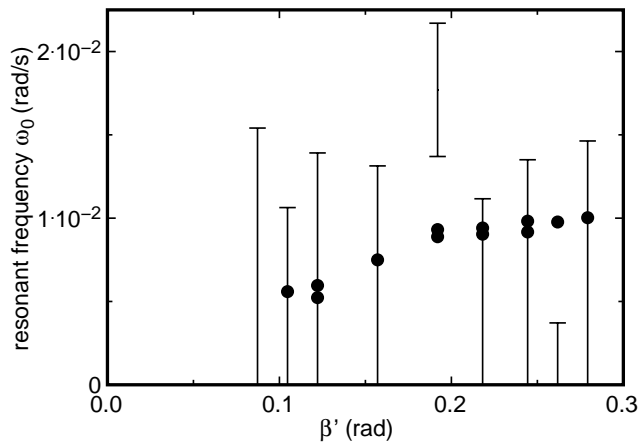}
\caption{The resonant angular frequency $\omega_0$ of tilt-induced planar oscillations in a cylindrical cell as a function of tilt angle $\beta'$.  Circles: measured values, reproduced from Ref.~\cite{BA08b}.  Error bars:  range of predicted $\omega_0$ using independently measured values $\omega_{\beta}^2$ and other parameters for each data point and propagating their errors. The model predicts the threshold value of $\beta'$ for resonance within 20\%, and the range of predicted resonance frequencies is consistent with the measured values.  This confirms that the effects of tilt on the LSC orientation are consistent with Eq.~\ref{eqn:ddottheta} and the values of $\omega_{\beta}^2$ obtained from $p(\theta_0)$.
 }
\label{fig:tilt_osc_revisited}
\end{figure} 

To resolve the discrepancy in the magnitude $\omega_{\beta}^2$  of the tilt-induced forcing on the dynamics of $\theta_0$, we revisit the prediction of tilt-induced oscillations, now including the damping term from Eq.~\ref{eqn:ddottheta} in the analysis and including the tilt-dependence of parameters.  Equation \ref{eqn:ddottheta} reduces to a damped driven linear harmonic oscillator when the potential is approximated as quadratic for small $\theta_0-\theta_{\beta}$, and if we assume fluctuations in $\delta$ are negligible so that the damping factor is $\bar\delta/\tau_{\dot\theta}\delta_0$.  The resonant angular frequency $\omega_0$ is given by:
\begin{equation}
\omega_0 = \sqrt{\omega_{\beta}^2\sin\beta' -\frac{1}{2}\left(\frac{\bar\delta}{\delta_0\tau_{\dot\theta}}\right)^2} \ .
\label{eqn:resonant_frequency_tilt}
\end{equation}
We compare the measured oscillation frequencies $\omega_0$ \cite{BA08b} with this model prediction to check for consistency of parameter values.  
We calculate $\omega_0$ from Eq.~\ref{eqn:resonant_frequency_tilt} using parameter values measured separately at each tilt angle.  $\tau_{\dot\theta}$ is obtained from the mean-squared displacement of $\dot\theta_0$ over time as the crossover between diffusive and plateau regimes \cite{BA08a}.  Because the value of $\omega_0$ is sensitive to parameter values near the resonance threshold, and there are large errors on $\omega_{\beta}^2$ dominated by the fit of Eq.~\ref{eqn:Deltalnptheta}, the best estimate for the term inside the square root of Eq.~\ref{eqn:resonant_frequency_tilt} is in many cases negative, while the errors often span positive and negative values. Thus, in Fig.~\ref{fig:tilt_osc_revisited} we plot the prediction as error bars covering this range representing $\pm1$ standard deviation, which extend down to zero for most of this data.  With these errors, the smallest $\beta'$ where we find parameter values consistent with resonance is 0.09 rad, just below the observed threshold of 0.10 rad.   We note that accounting for the effect of variable damping on the resonant frequency \cite{BA09} reduces the effective damping by no more than 3\% for the data where resonance is found in Fig.~\ref{fig:tilt_osc_revisited}, and the resulting shifts in predictions of $\omega_0$ are small compared to the errors.   At tilt angles $\beta'$ where resonance is found, the measured data is statistically consistent with the range of the predicted resonant frequencies, with a reduced $\chi^2=0.75$.  Thus, within these generous errors on average 60\% larger than the measured $\omega_0$, the model is consistent with the data.  However, the sensitivity of resonance to model parameters near the resonance threshold means the model doesn't accurately predict whether resonance occurs, similar to the case of oscillations around corners with restoring force due to the geometry of the cell \cite{JB20a}. The sensitivity of the resonance frequency to parameter values near the resonance threshold explains why erroneous values of $\omega_{\beta}$ were inferred from our previous work \cite{BA08b}, as the damping term shifts the resonance frequency downward significantly. 

 Despite the large errors in Fig.~\ref{fig:tilt_osc_revisited}, a chi-squared test narrows down the range of values of $\omega_{\beta}^2$ consistent with the data. In this case the uncertainty in the error squared is dominated by the error on the prediction of $\omega_{\beta}^2$ from the fit of Eq.~\ref{eqn:Deltalnptheta}.  To test the range of consistent values of $\omega_{\beta}^2$, we calculate a reduced $\chi^2$ with a prefactor in front of the measured $\omega_{\beta}^2$ in Eq.~\ref{eqn:resonant_frequency_tilt}.  Prefactors between 0.6 and 1.4 lead to a reduced $\chi^2<1.9$, which is the 95\% confidence interval for 8 data points.  This identifies the range of $\omega_{\beta}^2$ of $\pm40\%$ around the measured value yield predictions statistically consistent with the measured $\omega_0$.  If instead we fix $\omega_{\beta}^2$, values of $\omega_{\beta}^2 = 0.0038\pm0.0011$ s$^{-2}$ yield predictions consistent with the measured $\omega_0$ within a 95\% confidence interval.  This range is 20-40 times the previously predicted value of $\omega_{\beta}^2= 1.3\times10^{-4}$ s$^{-2}$ \cite{BA08b}.  The predicted value of $\omega_{\beta}^2= 1.3\times10^{-4}$ s$^{-2}$ results in a reduced $\chi^2=5.4$, inconsistent with the measured $\omega_0$ with a confidence interval better than 99.999\% for this model.

\section{Discussion: Possible causes of tilt sensitivity}
\label{sec:discussion}

The above analysis found a disagreement of about 2 orders of magnitude between the prediction and measurement of the tilt-induced-forcing parameter $\omega_{\beta}^2$. A model of diffusion in a potential self-consistently described $p(\theta_0)$ at different tilt angles, the change in the barrier-crossing rate with tilt for different barriers, and the frequency of tilt-induced oscillations without adjusting parameter values.  This confirms the functional form of Eq.~\ref{eqn:ddottheta} and the tilt-induced potential $V_{\beta}\propto -\cos(\theta_0-\theta_{\beta})$ are reliable for modeling such dynamics, and so these are not responsible for the disagreement between the measured and predicted values of $\omega_{\beta}^2$.   By process of elimination, this indicates the problem is with the predicted value of $\omega_{\beta}^2$.
  
The original prediction of the magnitude of $\omega_{\beta}^2$ assumed that the forcing only occurred in the thermal boundary layers, since that was all that was relevant for the forcing on $\delta$ which was the primary focus of that work \cite{BA08b}.  Here we additionally consider the  buoyant acceleration on the temperature difference $\delta$ in the bulk.  The buoyant acceleration near the sidewall is estimated to be $\dot u' = g\alpha\delta$.  The vector component of this that is parallel to the sidewall -- which is responsible for the term that drives the LSC -- is $\dot u'_{\phi} = g\alpha\delta\cos\beta\cos(\theta_0-\theta_\beta)$.  The $\beta$-dependence of this term was not even considered in the model \cite{BA08b}, since it only leads to a 4\% reduction in forcing (opposite the observed trend in $\bar\delta$), and it its tilt-dependence is weaker than the forcing on the thermal boundary layer \cite{BA08b}.   However, the corresponding component of the bulk forcing in the $\theta_0$-direction is $\dot u'_{\theta} = g\alpha\delta\sin\beta\sin(\theta_0-\theta_\beta)$ was also ignored in the original model \cite{BA08b}.  This enhances the forcing in $\theta_0$ significantly because of the stronger variation in $\sin\beta$ than $\cos\beta$ for small $\beta$.  The additional angular acceleration is estimated using the radius $L/2$ for a cylindrical cell to be
 $\ddot \theta_0' = 2\dot u'_{\theta}/L =  (2g\alpha\delta/L)\sin\beta\sin(\theta_0-\theta_\beta)$.   This yields $\omega_{\beta}^2 =2g\alpha\delta/L = 2.8\times10^{-3}$ s$^{-2}$ for the cubic cell.
 While this is nearly an order of magnitude larger than the boundary layer contribution predicted from \cite{BA08b}, it is still about an order of magnitude smaller than the measured value of $\omega_{\beta}^2 = 0.035$ s$^{-2}$.  
 
 The underprediction of the magnitude of tilt-induced forcing $\omega_{\beta}^2$, as well as the leveling off of the forcing $\omega_{\beta}^2\sin\beta'$ at large $\beta'$ in Fig.~\ref{fig:omega_beta2},  suggests there could be a relevant aspect of the temperature profile which is not captured by the model.  This could include changes to the shape of the LSC, counter-rolls, or other aspects of the temperature profile with tilt.  Such a feature would have to depend on both tilt and the LSC orientation $\theta_0$ relative to the tilt-orientation $\theta_{\beta}$ to affect $\omega_{\beta}^2$.

\subsection{Updated model prediction of asymmetric heating}

A model for the effect of asymmetric heating was also proposed previously \cite{BA08b}.  The extra forcing on the boundary layer was proposed to have magnitude $g\alpha\delta T/(\pi Nu L)  = 1.5\times 10^{-6}$ rad/s$^2$ where $\delta T = 0.005\Delta T/\sqrt{8}$ is the systematic horizontal temperature difference in the plates due to imperfections in the temperature control, for the cubic cell at $\Delta T = 18.4$ K and Nu = 82.8 \cite{JB20a}.  However, this prediction is much smaller than the measured effect, which was found to be somewhat larger than the geometric forcing with measured magnitude $\omega_r^2 = 2.2\times10^{-4}$ s$^{-2}$ \cite{JB20a}.  That prediction also ignored the forcing in the bulk.  Assuming the horizontal plate temperature difference $\delta T$ extends into the bulk, we instead approximate the additional bulk similar to the tilt-induced forcing as $ g\alpha\delta T/(\pi L) = 1.2\times10^{-4}$ s$^{-2}$, which is comparable to the geometric forcing, as observed \cite{JB20a}.


\section{Conclusions}

  The tilt-induced potential $V_{\beta}(\theta_0)$ acting on the LSC orientation $\theta_0$ in tilted Rayleigh-B{\'e}nard convection cells is obtained from measurements of the probability distribution $p(\theta_0)$.  The form of the potential $V_{\beta}(\theta_0$) is sinusoidal in $\theta_0-\theta_{\beta}$, and linear in tilt angle $\beta'$ for small $\beta'$, which is explained by a simple geometric model of the vector direction of the mean buoyancy force acting on the fluid. The magnitude of this tilt-induced forcing $\omega_{\beta}^2$ is used in a model of diffusion in a potential \cite{BA08b} to predict the change in barrier crossing rates for the LSC to escape different corners of a cubic cell as it is tilted, which are measured to be within 30\% of predictions.   The changes in barrier-crossing rate with tilt are due to tilt-induced changes in the heights of the potential barriers that separate corners of the cubic cell, as well as changes in the height of the local minima of the potential, which affects the probability of the LSC being in each well.  The model also predicts tilt-induced oscillations as the tilt-induced potential provides enough of a restoring force to overcome damping and produce resonance in a cylindrical cell at large tilt angles.  The smallest tilt angle where these oscillations are found is predicted within 20\% of measurements, and predicted oscillation frequencies are consistent with measurements, when using the same value of $\omega_{\beta}^2$ as obtained from $p(\theta_0)$.  These observations show that a self-consistent model of diffusion in a potential \cite{BA08b} characterizes the dynamics of the LSC orientation $\theta_0$ in tilted cells.

However, a previous prediction of the magnitude of the tilt-induced forcing $\omega_{\beta}^2$  \cite{BA08b} is about two orders of magnitude smaller than the measurement.  Including the buoyant forcing on the bulk of the LSC -- which was ignored in the original prediction that considered the forcing on the thermal boundary layers only \cite{BA08b} -- reduces the difference to one order of magnitude.  The failure of the model to predict the magnitude of the tilt-induced forcing remains an open issue.  The forcing due to tilt is also found to reach a plateau at large tilt angles instead of following the linear prediction, another failure of the prediction which remains an open issue.

\section{Acknowledgments}

We thank the University of California, Santa Barbara machine shop and K. Faysal for helping with construction of the experimental apparatus. This work is supported by Grant No.~CBET-1255541 of the U.S. National Science Foundation.  

Experiments in cubic cells were carried out by D. Ji. Preliminary analysis identifying the disagreement between the prediction and measurements of the magnitude of the effect of tilt on the LSC orientation was carried out by K. Bai.  Analysis for Figs.~\ref{fig:lnptheta_cube} and \ref{fig:omega_beta} was carried out by D. Ji.  Analysis for Figs.~\ref{fig:omega_beta2} and \ref{fig:tilt_osc_revisited} was carried out by E. Brown.  Predictions of the tilt-modified barrier-crossing rate were carried out by E. Brown.  The article was written by E. Brown.

\section{Credit Line}

This article may be downloaded for personal use (i.e.~not for-profit use) only. Any other use requires prior permission of the author and AIP Publishing. This article appeared in Physics of Fluids 32 (7), 075118 (2020), and may be downloaded at:  https://doi.org/10.1063/5.0018051

\section{Data Availability}

The data and analysis that support the findings of this study are available from the corresponding author upon reasonable request.


%

\end{document}